\documentclass[reprint,twocolumn,superscriptaddress,amsmath,amssymb,showpacs,prl]{revtex4-1}
\usepackage{hyperref}
\pdfoutput=1 

\usepackage{dcolumn}
\usepackage{epsfig}
\usepackage{graphics}
\usepackage[latin1]{inputenc} 
\usepackage[T1]{fontenc}
\usepackage{bm}
\newcommand{\ddst}{false}

\begin{document}

\title{Stretched Exponential Relaxation of Glasses at Low Temperature}

 \author{Yingtian Yu}
 \affiliation{Physics of AmoRphous and Inorganic Solids Laboratory (PARISlab), Department of Civil and Environmental Engineering, University of California, Los Angeles, CA, USA}
 \author{Mengyi Wang}
 \affiliation{Physics of AmoRphous and Inorganic Solids Laboratory (PARISlab), Department of Civil and Environmental Engineering, University of California, Los Angeles, CA, USA}
 \author{Dawei Zhang}
 \affiliation{Physics of AmoRphous and Inorganic Solids Laboratory (PARISlab), Department of Civil and Environmental Engineering, University of California, Los Angeles, CA, USA}
 \author{Bu Wang}
 \affiliation{Physics of AmoRphous and Inorganic Solids Laboratory (PARISlab), Department of Civil and Environmental Engineering, University of California, Los Angeles, CA, USA}
\author{Gaurav Sant}
 \affiliation{Laboratory for the Chemistry of Construction Materials (LC$^2$), Department of Civil and Environmental Engineering, University of California, Los Angeles, CA, USA}
 \affiliation{California Nanosystems Institute (CNSI), University of California, Los Angeles, CA, USA}
\author{Mathieu Bauchy}
 \email[Contact: ]{bauchy@ucla.edu}
 \homepage[\\Homepage: ]{http://mathieu.bauchy.com}
 \affiliation{Physics of AmoRphous and Inorganic Solids Laboratory (PARISlab), Department of Civil and Environmental Engineering, University of California, Los Angeles, CA, USA}
 
\date{\today}

\pacs{65.60.+a, 62.40.+i, 81.05.Kf}

\begin{abstract}
The question of whether glass continues to relax at low temperature is of fundamental and practical interest. Here, we report a novel atomistic simulation method allowing us to directly access the long-term dynamics of glass relaxation at room temperature. We find that the potential energy relaxation follows a stretched exponential decay, with a stretching exponent $\beta = 3/5$,  as predicted by Phillips' diffusion-trap model. Interestingly, volume relaxation is also found. However, it is not correlated to the energy relaxation, but is rather a manifestation of the mixed alkali effect.
\end{abstract}

\maketitle

While it is indeed commonly believed that, as frozen supercooled liquids, glasses should continue to flow over the years (e.g. in the case of the stained-glass windows of medieval cathedrals \cite{zanotto_cathedral_1998, zanotto_cathedral_1999}), the dramatic increase of their viscosity below the glass transition temperature $T_{\rm g}$ suggests, on the contrary, that their relaxation time is on the order of 10$^{32}$ years at room temperature \cite{stokes_flowing_1999}. However, a recent study conducted by Mauro \textit{et al.} \cite{welch_dynamics_2013} reported the intriguing dynamics of the relaxation of a commercial Corning$^{\circledR}$ Gorilla$^{\circledR}$ Glass at room temperature, over 1.5 years. Fused silica \cite{vannoni_long-term_2010} and metallic glasses \cite{sahu_room_2009} were also reported to show an appreciable relaxation over time. More generally, the relaxation of glasses at low temperature is known as the "thermometer effect", originating from the fact that, in the 19$^{\rm st}$ century, thermometers made of alkali lime silicate glass were experiencing gradual changes of dimension over time, making them inaccurate \cite{kurkjian_perspectives_1998, bunde_ionic_1998}.

As it cannot be described as a viscous process, the physical origin of such room temperature relaxation remains unclear. However, the shape of the relaxation function contains valuable information to discriminate between the different proposed models. In particular, Mauro \textit{et al.} observed that the volume of the relaxing Gorilla$^{\circledR}$ Glass followed a stretched exponential decay function \cite{welch_dynamics_2013}, with a stretching exponent $\beta = 3/7$, which, interestingly, is the value predicted by the Phillips diffusion-trap model for a relaxation dominated by long-range pathways \cite{phillips_stretched_1996}. Understanding and predicting the relaxation dynamics of glasses has important practical applications, e.g. for optical fibers \cite{yang_effect_2013},  substrate glass for liquid crystal displays \cite{mauro_nonequilibrium_2009}, and chemically strengthened cover glass for smartphones and tablets \cite{tandia_atomistic_2012}.

Here, based on molecular dynamics (MD) simulations, we investigate the atomistic origin of the relaxation of realistic alkali aluminosilicate glasses at temperature far below $T_{\rm g}$. As traditional MD is unable to reproduce year-long relaxations, we introduce a novel artificial relaxation method based on cyclic stress  perturbations. The potential energy of the simulated glasses is found to follow a stretched-exponential decay function, with a stretching exponent $\beta = 3/5$, characteristic of systems dominated by short-range forces \cite{phillips_stretched_1996}. On the contrary, the volume relaxation appears composition-specific and strongly correlated with the dynamics of the alkali atoms.

For this study, we simulated a 2991 atoms mixed potassium sodium aluminosilicate glass (denoted KNAS hereafter), of composition (K$_2$O)$_8$(Na$_2$O)$_8$(Al$_2$O$_3$)$_9$(SiO$_2$)$_{75}$, which we expect to be representative of that of the peralkaline Corning$^{\circledR}$ Gorilla$^{\circledR}$ Glass used in Ref. \cite{welch_dynamics_2013}. In addition, in order to evaluate the effect of the mixed alkali effect on relaxation, we simulated a similar single alkali glass (denoted NAS hereafter) of composition (Na$_2$O)$_{16}$(Al$_2$O$_3$)$_9$(SiO$_2$)$_{75}$. All MD simulations are performed with the LAMMPS package \cite{plimpton_fast_1995}, using the well-established Teter potential \cite{cormack_alkali_2002, bauchy_structural_2012, bauchy_viscosity_2013, xiang_structure_2013} and an integration time-step of 1 fs. Coulomb interactions were evaluated by the Ewald summation method, with a cutoff of 12 \AA. The short-range interaction cutoff was chosen at 8.0 \AA. Liquids were first generated by placing the atoms randomly in the simulation box. The liquids were then equilibrated at 5000 K in the NPT ensemble  (constant pressure) for 1 ns, at zero pressure, to assure the loss of the memory of the initial configuration. Glasses were formed by linear cooling of the liquids from 5000 to 300 K with a cooling rate of 1 K/ps in the NPT ensemble at zero pressure. The obtained glasses were eventually relaxed for an additional 1 ns in the NPT ensemble at 300 K and zero pressure. A detailed structural analysis of the simulated glasses can be found in Ref. \cite{xiang_structure_2013}.

Traditional MD simulations are usually limited to a few nanoseconds, which prevents one from using them to predict long-term relaxation at low temperature. On the other hand, kinetic Monte-Carlo simulations like ART \cite{barkema_event-based_1996} offer an attractive alternative to perform simulations up to a few seconds, but their application to alkali silicate is challenging due to the high mobility of the alkali atoms, which results in a huge number of small energy barriers to compute. Since a direct simulation of relaxation dynamics is, at this point, unachievable, we developed a new method to simulate an artificial relaxation.

\begin{figure}
\begin{center}
\includegraphics*[width=\linewidth, keepaspectratio=true, draft=\ddst]{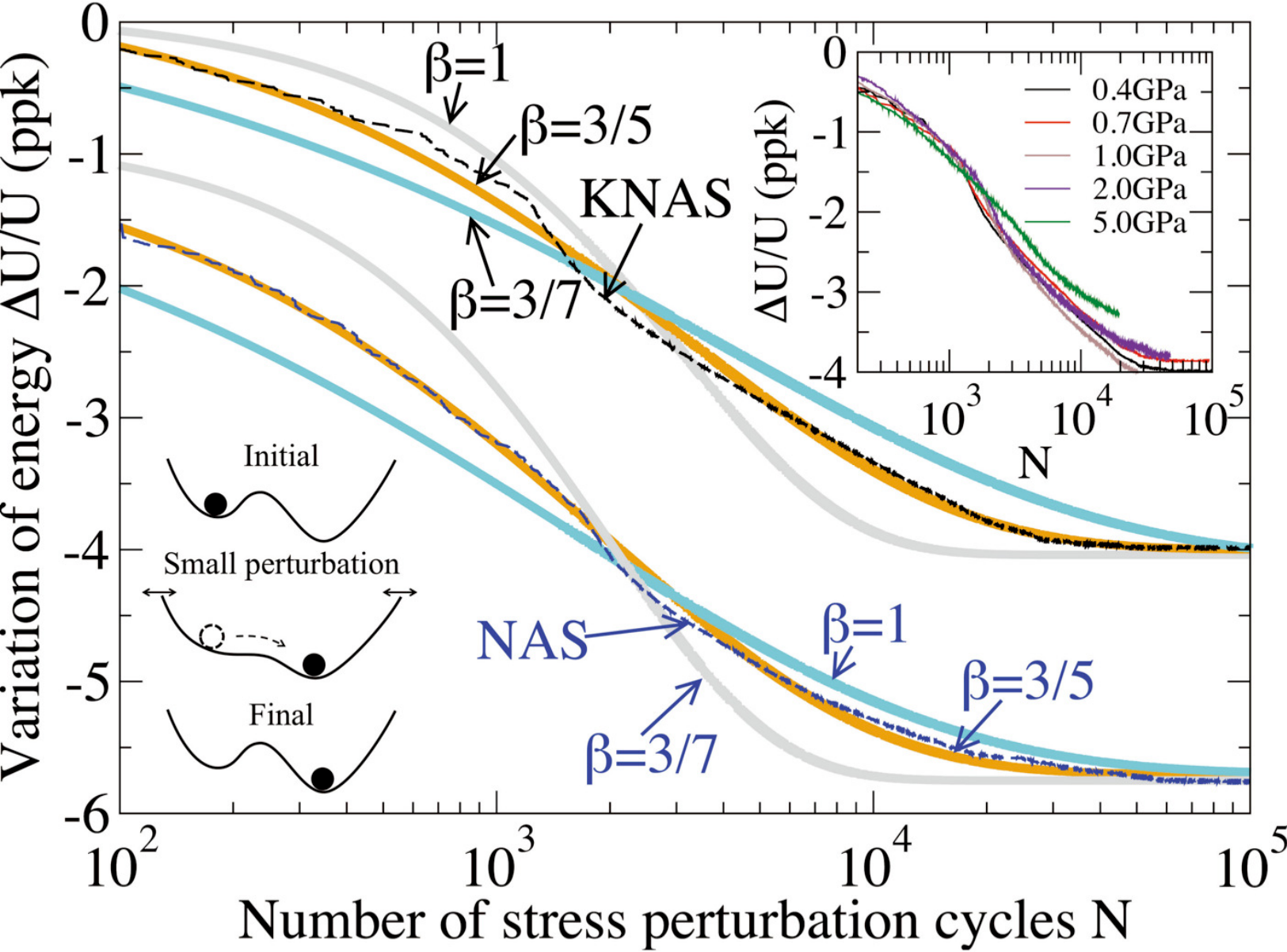}
\caption{\label{fig:energy} (color online). Relative variation of the potential energy (stabilization) of the simulated mixed potassium sodium aluminosilicate (KNAS) and sodium aluminosilicate (NAS) glasses, with respect to the number of stress perturbation cycles applied $N$ (with an stress amplitude of 0.4 GPa). For better clarity, the curve obtained for the NAS glass is shifted by $-1$. The simulated data are fitted by stretched exponential decay functions with different stretching exponents $\beta$ (solid lines). The inset shows the relative variation of the potential energy of the simulated KNAS glass, for different amplitudes of stress perturbations. The cartoon on the bottom-left is a schematic representartion of a strain-induced disappearance of energy barrier. The W curves represent the potential energy between two meta-stable equilibrium states, and the circles represent the state of the system. The arrow corresponds to a strain-activated jump of the system towards the more stable equilibrium state.
}
\end{center}	
\end{figure}

This method is inspired by the artificial aging or rejuvenation observed in granular materials subjected to vibrations \cite{richard_slow_2005, mobius_irreversibility_2014}. Small vibrations induce a compaction of the material, that is, they make the system artificially age. On the other hand, large vibrations randomize the grain arrangements, which decreases the overall compactness and, therefore, make the system rejuvenate. Similar ideas, relying on the energy landscape approach \cite{lacks_energy_2001, lacks_energy_2004}, have been applied to amorphous solids, based on the fact that small stresses deform the energy landscape undergone by the atoms. This can result in the removal of some energy barriers existing at zero stress, thus allowing atoms to jump over them in order to relax to lower energy states (see the schematic representation in Fig. \ref{fig:energy}). This transformation is irreversible as, once the stress is removed, the system remains in its "aged" state. On the contrary, large stresses move the system far from its initial state, which eventually leads to rejuvenation, similar to thermal annealing \cite{utz_atomistic_2000, lyulin_time_2007}. As such, several experimental and simulation studies reported that glasses can undergo overaging or rejuvenation, when subjected to small or large shear stresses, respectively \cite{rottler_deformation_2008, bonn_laponite:_2002, viasnoff_rejuvenation_2002, utz_atomistic_2000, priezjev_heterogeneous_2013, lacks_energy_2001, lacks_energy_2004, fiocco_oscillatory_2013, lyulin_time_2007}. The novel method presented here is based on a succession of many of such stress perturbations, inducing successive small relaxations of the system.

In practice, a glass initially relaxed at zero pressure is subjected to athermal stress perturbations cycles. Each perturbation cycle consists, successively, of (1) an hydrostatic compressive stress $-\sigma_0$ and (2) an hydrostatic tensile stress $\sigma_0$. Therefore, the average stress remains zero over each cycle. During both of the phases (1) and (2), a volume-variant energy minimization is performed, in which the positions of the atoms and the volume/shape of the simulation box are updated in an iterative process, in order for the system to achieve simultaneously both an energy minimum for the potential energy of the atoms $U$ and a stress $\sigma$ close to $\pm \sigma_0$. This combined optimization is performed by minimizing a fictitious energy $E_{\rm fic} = U + E_{\rm strain}$, where $U$ is the potential energy and $E_{\rm strain}$ a strain energy expression proposed by Parrinello and Rahman \cite{parrinello_polymorphic_1981}. Hence, during each cycle, the system is allowed to jump over energy barriers, with a roughly constant activation energy on the order of $E_{\rm strain}$. According to transition state theory \cite{vineyard_frequency_1957}, each cycle should roughly correspond to a constant time of about $\Delta t = \nu_0^{-1} \exp(E_{\rm strain}/k_{\rm B}T)$, where $\nu_0$ is a characteristic frequency of attempt, $k_{\rm B}$ the Boltzmann constant, and $T$ the temperature \cite{masoero_kinetic_2013}. Therefore, the relaxation observed with respect to the number of stress perturbations cycles applied $N$ should be representative of the relaxation versus time $t$, with a converting factor $t=N \Delta t$.

\begin{figure}
\begin{center}
\includegraphics*[width=\linewidth, keepaspectratio=true, draft=\ddst]{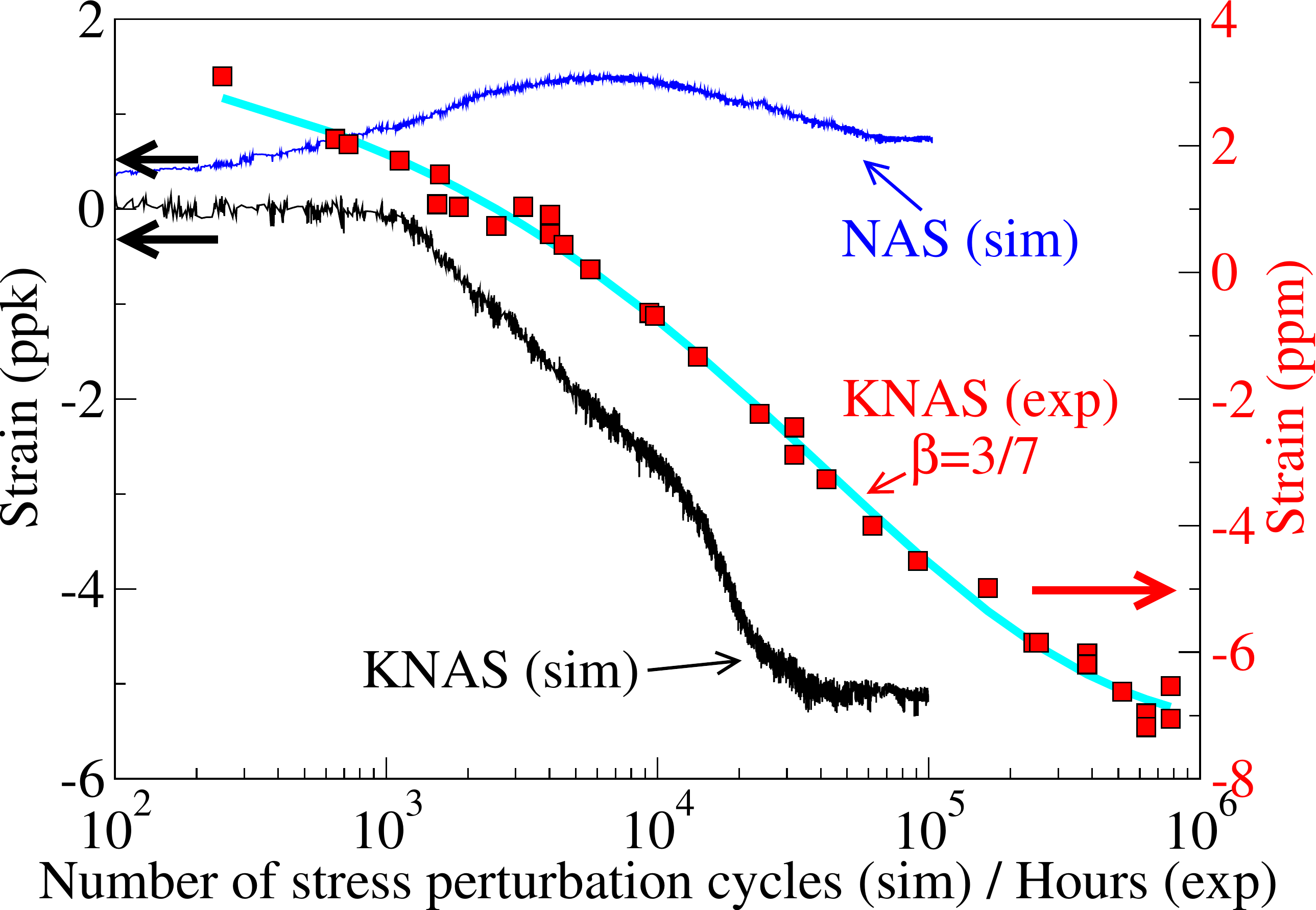}
\caption{\label{fig:volume} (color online). Linear strain (compression when negative) observed during the low temperature relaxation of the simulated mixed potassium sodium aluminosilicate (KNAS) and sodium aluminosilicate (NAS) glasses (left axis) with respect to the number of stress perturbation cycles applied $N$. Simulated results are compared to the experimental measurement of the room temperature compaction of a commercial KNAS glass (right axis) with respect to the time of relaxation \cite{welch_dynamics_2013}. Experimental data are fitted by a stretched exponential decay function (solid line) with a stretching exponent $\beta = 3/7$.
}
\end{center}	
\end{figure}


Figure \ref{fig:energy} shows the relative variation of the potential energies of the KNAS and NAS glasses, with respect to the number $N$ of stress perturbation cycles applied, with a sub-yield amplitude $\sigma_0 = 0.4$ GPa. As expected, the stress perturbations allow the system to relax towards lower energy states. This stabilization is gradual and about $10^5$ of such cycles are needed for the potential energy to plateau. Eventually, a significant decrease of energy (around $0.4 \%$) is achieved. We note that the energy relaxation profile appears to be very similar for the KNAS and NAS glasses. The influence of $\sigma_0$ was assessed by performing similar simulations with $\sigma_0 =$ 0.7, 1, 2, and 5 GPa. As shown in the inset of Fig. \ref{fig:energy}, no significant change of the shape of the energy relaxation is observed at low $\sigma_0$. Therefore, the relaxation dynamics does not depend on the choice of $\sigma_0$. However, the energy starts to decrease more slowly for $\sigma_0 = 5$ GPa, which suggests that, for such a high stress, the system is slightly rejuvenated at each cycle. In the following, we keep $\sigma_0 = 0.4$ GPa. Overall, this result constitutes the first direct simulation of the relaxation of a glass at low temperature, to the best of our knowledge.


Interestingly, the volume of the glasses does not remain constant through the relaxation procedure. As shown in Fig. \ref{fig:volume}, the simulated KNAS glass shows a gradual compaction with respect to the number of stress perturbation cycles applied, eventually achieving a final linear strain of about $-0.5\%$. As the average applied stress remains zero, this compaction cannot be explained by elastic deformations. We note that the extent of the compaction is higher in the present simulation than observed experimentally \cite{welch_dynamics_2013}. This discrepancy is likely to arise from the cooling rate used in simulation, which is much higher than the ones typically achieved experimentally, and from the different compositions of the glasses. However, as shown in Fig. \ref{fig:volume}, the general trend of the volume relaxation is remarkably fairly similar to that observed experimentally.

The origin of the densification of the Gorilla$^{\circledR}$ Glass and, more generally, of the thermometer effect, has been attributed to the mixed alkali effect (MAE) \cite{bunde_ionic_1998, kurkjian_perspectives_1998}. An MAE is generally observed in silicate or borate glasses containing at least two different network modifying alkali oxides AO$_2$ and BO$_2$. It manifests by a strong non linear evolution of the transport properties with respect to the fraction A/B. Here, the compaction of the KNAS glass can be clearly attributed to the MAE. Indeed, as shown in Fig. \ref{fig:volume}, the volume of the single alkali NAS glass features a completely different trend, remaining fairly constant over time. This result is consistent with the observation that single alkali-containing Gorilla$^{\circledR}$ Glass 2 and 3 do not show any discernable volume relaxation \cite{welch_dynamics_2013}. These results highlight the fact that volume relaxation is not an intrinsic feature of glass, but strongly depends on composition. On the contrary, based on the present results, energy relaxation appears to be more general and, therefore, more relevant to characterize the intrinsic dynamics of glass relaxation.

\begin{figure}
\begin{center}
\includegraphics*[width=\linewidth, keepaspectratio=true, draft=\ddst]{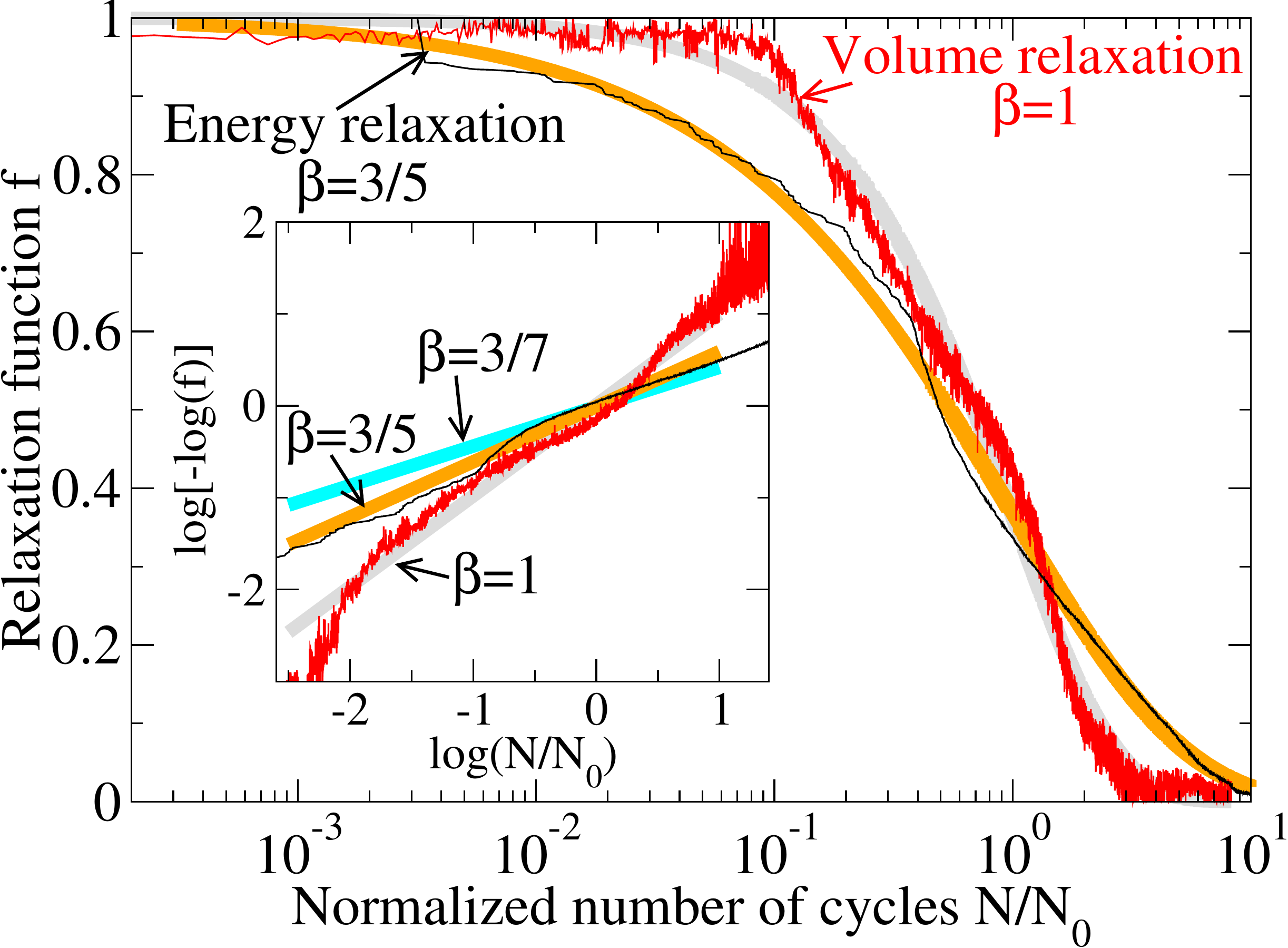}
\caption{\label{fig:relax} (color online). Relaxation function $f$ for the energy and volume, computed during the relaxation of the simulated mixed potassium sodium aluminosilicate (KNAS) glass, with respect to the normalized number of stress perturbation cycles applied $N/N_0$ (see text). The computed data are fitted with stretched exponential decay functions, with stretching exponents $\beta = 3/5$ and $1$ for the energy and the volume, respectively. The inset shows $\log(-\log(f))$ with respect to $\log(N/N_0)$, for the energy and volume, whose slope is equal to $\beta$. Solid lines with the slopes $\beta=3/7$, $3/5$, and $1$ are added for comparison.
}
\end{center}	
\end{figure}


Indeed, the general shape of the relaxation curve of the potential energy is of fundamental importance as it manifests from the topology of the relaxation process \cite{potuzak_topological_2011}. Glass relaxation behaviors are generally found to follow a stretched exponential decay function $f(t)$ \cite{simdyankin_relationship_2003}:

\begin{equation}
 f(t) = \exp(-(t/\tau)^{\beta})
\end{equation}
where $\tau$ is the relaxation time, and $\beta$ a dimensionless stretching exponent satisfying $0 < \beta < 1$. The case $\beta = 1$ corresponds to a simple exponential decay. The diffusion-trap model from Phillips \cite{phillips_stretched_1996}, based on diffusion of excitations to randomly distributed traps, predicts that a theoretical value for the stretching exponent $\beta = d^\ast/(d^\ast + 2)$, where $d^\ast$ is the effective dimensionality of the channels along which the excitations diffuse in the configuration space. This is described as $d^\ast = f d$, where $d$ is the dimensionality of the system ($3$ for structural glasses), and $f$ is the proportion of the active channels of relaxation. Hence, when all the channels are active ($f=1$), one gets $\beta = 3/5$. When only long-range channels are active, by assuming an equipartitioning of the short- and long-range contributions ($f =1/2$), the model predicts $\beta = 3/7$.

As shown in Fig. \ref{fig:volume}, the experimental volume relaxation of the Gorilla$^{\circledR}$ Glass was found to follow such a stretched exponential decay function, with a relaxation time $\tau = 27.6$ days and a stretching exponent $\beta = 3/7$, attributed to a relaxation dominated by long-range pathways. Interestingly, the energy relaxation predicted by the present simulations also features a stretched exponential decay, with an equivalent relaxation time $\tau = N_0 \Delta t$, with $N_0 = 3200$ and $1300$ for the KNAS and NAS glasses, respectively. This allows us to roughly evaluate $\Delta t$, the fictitious time elapsed during each stress perturbation cycle, to be on the order of $10$ min. However, as shown in Fig. \ref{fig:energy}, we unambiguously find a stretching exponent $\beta = 3/5$, which, as mentioned previously, corresponds to the situation in which all the relaxation channels are active \cite{phillips_stretched_1996}, which is typically observed for the stress relaxation in glasses \cite{potuzak_topological_2011, mauro_unified_2012}. We note that the $\beta = 3/5$ factor offers a good fit, both for the KNAS and NAS glasses, which suggests that it does not depend on changes of composition.

Surprisingly, as shown in Fig. \ref{fig:relax}, the energy and volume relaxations of the KNAS glass do not show the same shape. First, we find $N_0 = 3200$ and $12000$ for the energy and volume, respectively, suggesting a slower relaxation for the volume than for the energy. Second, the volume relaxation actually follows a simple (that is, not stretched) relaxation decay function. As shown in the inset of Fig. \ref{fig:relax}, this difference of stretching exponent can be clearly established by plotting $\log(-\log(f))$ versus $\log(N/N_0)$, whose slope is equal to $\beta$. This result suggests that energy and volume relax in a fundamentally different way. More specifically, the relaxation of energy can be explained by the diffusion-trap model with all relaxation channels being active \cite{phillips_stretched_1996}, whereas the relaxation of volume, showing a simple exponential relaxation, can be explained by a simple two-state model \cite{mauro_minimalist_2012}.

\begin{figure}
\begin{center}
\includegraphics*[width=\linewidth, keepaspectratio=true, draft=\ddst]{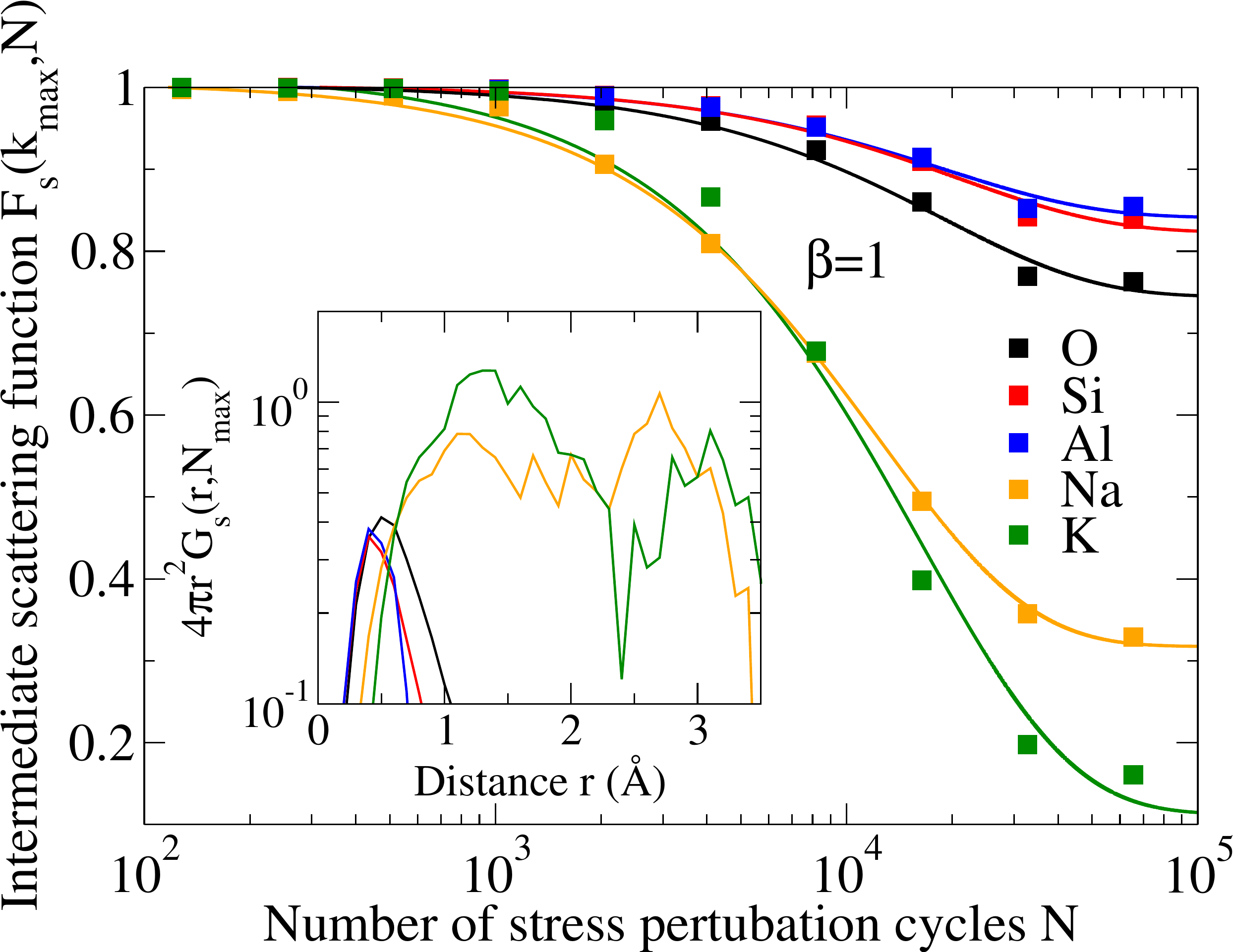}
\caption{\label{fig:isf} (color online). O, Si, Al, Na, and K intermediate scattering function $F_{\rm s}(k,N)$ (computed at $k_{\rm max}$, the position of the first peak of the structure factor) of the simulated potassium sodium aluminosilicate (KNAS) glass with respect to the number of stress perturbation cycles applied $N$. The simulated data are fitted by simple exponential decay functions ($\beta = 1$, solid lines). The inset shows the self-part of the van Hove correlation function $4 \pi r^2 G_{\rm s}(r,N)$ for O, Si, Al, Na, and K, with respect to the distance $r$, computed at $N_{\rm max} = 2^{16}$ cycles.
}
\end{center}	
\end{figure}


In order to elucidate the origin of the simple exponential decay for the volume, we investigated the relaxation dynamics of each atomic species by calculating the intermediate scattering function (ISF) $F_{\rm s}(k,N)$, computed at $k_{\rm max}$, the position of the first peak of the structure factor \cite{bauchy_densified_2015}. Interestingly, the ISF of Na and K atoms follow a trend similar to that of the volume relaxation. Indeed, as shown in Fig. \ref{fig:isf}, it can be fitted by a simple exponential decay function ($\beta = 1$) and show relaxation times similar to that of the volume relaxation (e.g., $N_0 = 11900$ for Na atoms, as compared to $12000$ for the volume relaxation). This clearly establishes that the volume relaxation observed in mixed alkali aluminosilicate glasses arises from the relaxation dynamics of the alkali atoms. The self part of the van Hove correlation function, shown in the inset of Fig. \ref{fig:isf}, shows that this relaxation of the alkali atoms consists of jumps, with a typical hopping distance of around 3 \AA\ \cite{bauchy_pockets_2011, abdolhosseini_qomi_anomalous_2014, mousseau_cooperative_2000}. On the contrary, the network former atoms show very limited motion, lower than 1 \AA. This highlights the fact that the volume relaxation is not correlated to the viscosity of the glass network, but rather to the dynamics of the alkali modifiers only. The fact that, due to the MAE effect, this dynamics strongly depends on the composition of the glass may explain the discrepancy of stretching exponent between the simulation and experimental results.

Featuring a stressed exponential decay function with $\beta = 3/5$, the shape of the energy relaxation is, according to the present results, more intrinsic to the glassy state than that of the volume relaxation. However, if the shape appears universal, the typical relaxation time $\tau$ strongly depends on composition. In particular, relaxation has been shown to be fastest for isostatic systems, which are characterized by an atomic network rigid but free of internal stress \cite{bauchy_densified_2015, chakravarty_ageing_2005}. Beyond glasses, being able to predict and tune the relaxation and aging of materials could improve the understanding of memory encoded materials \cite{fiocco_encoding_2014, fiocco_memory_2015} or protein folding \cite{mousseau_sampling_2001, phillips_scaling_2009}.

\begin{acknowledgments}

The authors acknowledge financial support for this research provisioned by the University of California, Los Angeles (UCLA). GNS acknowledges the National Science Foundation (CAREER: 1253269) for partial support of this work.
\end{acknowledgments}


\begin{thebibliography}{45}
\expandafter\ifx\csname natexlab\endcsname\relax\def\natexlab#1{#1}\fi
\expandafter\ifx\csname bibnamefont\endcsname\relax
  \def\bibnamefont#1{#1}\fi
\expandafter\ifx\csname bibfnamefont\endcsname\relax
  \def\bibfnamefont#1{#1}\fi
\expandafter\ifx\csname citenamefont\endcsname\relax
  \def\citenamefont#1{#1}\fi
\expandafter\ifx\csname url\endcsname\relax
  \def\url#1{\texttt{#1}}\fi
\expandafter\ifx\csname urlprefix\endcsname\relax\def\urlprefix{URL }\fi
\providecommand{\bibinfo}[2]{#2}
\providecommand{\eprint}[2][]{\url{#2}}

\bibitem[{\citenamefont{Zanotto}(1998)}]{zanotto_cathedral_1998}
\bibinfo{author}{\bibfnamefont{E.~D.} \bibnamefont{Zanotto}},
  \bibinfo{journal}{American Journal of Physics} \textbf{\bibinfo{volume}{66}},
  \bibinfo{pages}{392} (\bibinfo{year}{1998}).

\bibitem[{\citenamefont{Zanotto and Gupta}(1999)}]{zanotto_cathedral_1999}
\bibinfo{author}{\bibfnamefont{E.~D.} \bibnamefont{Zanotto}} \bibnamefont{and}
  \bibinfo{author}{\bibfnamefont{P.~K.} \bibnamefont{Gupta}},
  \bibinfo{journal}{American Journal of Physics} \textbf{\bibinfo{volume}{67}},
  \bibinfo{pages}{260} (\bibinfo{year}{1999}).

\bibitem[{\citenamefont{Stokes}(1999)}]{stokes_flowing_1999}
\bibinfo{author}{\bibfnamefont{Y.~M.} \bibnamefont{Stokes}},
  \bibinfo{journal}{Proceedings of the Royal Society of London A: Mathematical,
  Physical and Engineering Sciences} \textbf{\bibinfo{volume}{455}},
  \bibinfo{pages}{2751} (\bibinfo{year}{1999}).

\bibitem[{\citenamefont{Welch et~al.}(2013)\citenamefont{Welch, Smith, Potuzak,
  Guo, Bowden, Kiczenski, Allan, King, Ellison, and
  Mauro}}]{welch_dynamics_2013}
\bibinfo{author}{\bibfnamefont{R.}~\bibnamefont{Welch}},
  \bibinfo{author}{\bibfnamefont{J.}~\bibnamefont{Smith}},
  \bibinfo{author}{\bibfnamefont{M.}~\bibnamefont{Potuzak}},
  \bibinfo{author}{\bibfnamefont{X.}~\bibnamefont{Guo}},
  \bibinfo{author}{\bibfnamefont{B.}~\bibnamefont{Bowden}},
  \bibinfo{author}{\bibfnamefont{T.}~\bibnamefont{Kiczenski}},
  \bibinfo{author}{\bibfnamefont{D.}~\bibnamefont{Allan}},
  \bibinfo{author}{\bibfnamefont{E.}~\bibnamefont{King}},
  \bibinfo{author}{\bibfnamefont{A.}~\bibnamefont{Ellison}}, \bibnamefont{and}
  \bibinfo{author}{\bibfnamefont{J.}~\bibnamefont{Mauro}},
  \bibinfo{journal}{Physical Review Letters} \textbf{\bibinfo{volume}{110}},
  \bibinfo{pages}{265901} (\bibinfo{year}{2013}).

\bibitem[{\citenamefont{Vannoni et~al.}(2010)\citenamefont{Vannoni, Sordini,
  and Molesini}}]{vannoni_long-term_2010}
\bibinfo{author}{\bibfnamefont{M.}~\bibnamefont{Vannoni}},
  \bibinfo{author}{\bibfnamefont{A.}~\bibnamefont{Sordini}}, \bibnamefont{and}
  \bibinfo{author}{\bibfnamefont{G.}~\bibnamefont{Molesini}},
  \bibinfo{journal}{Optics Express} \textbf{\bibinfo{volume}{18}},
  \bibinfo{pages}{5114} (\bibinfo{year}{2010}).

\bibitem[{\citenamefont{Sahu et~al.}(2009)\citenamefont{Sahu, Gangopadhyay,
  Kelton, Chatterjee, and Sahoo}}]{sahu_room_2009}
\bibinfo{author}{\bibfnamefont{R.}~\bibnamefont{Sahu}},
  \bibinfo{author}{\bibfnamefont{A.~K.} \bibnamefont{Gangopadhyay}},
  \bibinfo{author}{\bibfnamefont{K.~F.} \bibnamefont{Kelton}},
  \bibinfo{author}{\bibfnamefont{S.}~\bibnamefont{Chatterjee}},
  \bibnamefont{and} \bibinfo{author}{\bibfnamefont{K.~L.} \bibnamefont{Sahoo}},
  \bibinfo{journal}{Scripta Materialia} \textbf{\bibinfo{volume}{61}},
  \bibinfo{pages}{588} (\bibinfo{year}{2009}).

\bibitem[{\citenamefont{Kurkjian and
  Prindle}(1998)}]{kurkjian_perspectives_1998}
\bibinfo{author}{\bibfnamefont{C.~R.} \bibnamefont{Kurkjian}} \bibnamefont{and}
  \bibinfo{author}{\bibfnamefont{W.~R.} \bibnamefont{Prindle}},
  \bibinfo{journal}{Journal of the American Ceramic Society}
  \textbf{\bibinfo{volume}{81}}, \bibinfo{pages}{795} (\bibinfo{year}{1998}).

\bibitem[{\citenamefont{Bunde et~al.}(1998)\citenamefont{Bunde, Funke, and
  Ingram}}]{bunde_ionic_1998}
\bibinfo{author}{\bibfnamefont{A.}~\bibnamefont{Bunde}},
  \bibinfo{author}{\bibfnamefont{K.}~\bibnamefont{Funke}}, \bibnamefont{and}
  \bibinfo{author}{\bibfnamefont{M.~D.} \bibnamefont{Ingram}},
  \bibinfo{journal}{Solid State Ionics} \textbf{\bibinfo{volume}{105}},
  \bibinfo{pages}{1} (\bibinfo{year}{1998}).

\bibitem[{\citenamefont{Phillips}(1996)}]{phillips_stretched_1996}
\bibinfo{author}{\bibfnamefont{J.~C.} \bibnamefont{Phillips}},
  \bibinfo{journal}{Reports on Progress in Physics}
  \textbf{\bibinfo{volume}{59}}, \bibinfo{pages}{1133} (\bibinfo{year}{1996}).

\bibitem[{\citenamefont{Yang et~al.}(2013)\citenamefont{Yang, Sangleboeuf,
  Boussard-Pl{\'e}del, and Bureau}}]{yang_effect_2013}
\bibinfo{author}{\bibfnamefont{G.}~\bibnamefont{Yang}},
  \bibinfo{author}{\bibfnamefont{J.-C.} \bibnamefont{Sangleboeuf}},
  \bibinfo{author}{\bibfnamefont{C.}~\bibnamefont{Boussard-Pl{\'e}del}},
  \bibnamefont{and} \bibinfo{author}{\bibfnamefont{B.}~\bibnamefont{Bureau}},
  \bibinfo{journal}{Journal of the American Ceramic Society}
  \textbf{\bibinfo{volume}{96}}, \bibinfo{pages}{464} (\bibinfo{year}{2013}).

\bibitem[{\citenamefont{Mauro et~al.}(2009)\citenamefont{Mauro, Allan, and
  Potuzak}}]{mauro_nonequilibrium_2009}
\bibinfo{author}{\bibfnamefont{J.~C.} \bibnamefont{Mauro}},
  \bibinfo{author}{\bibfnamefont{D.~C.} \bibnamefont{Allan}}, \bibnamefont{and}
  \bibinfo{author}{\bibfnamefont{M.}~\bibnamefont{Potuzak}},
  \bibinfo{journal}{Physical Review B} \textbf{\bibinfo{volume}{80}},
  \bibinfo{pages}{094204} (\bibinfo{year}{2009}).

\bibitem[{\citenamefont{Tandia et~al.}(2012)\citenamefont{Tandia, Vargheese,
  Mauro, and Varshneya}}]{tandia_atomistic_2012}
\bibinfo{author}{\bibfnamefont{A.}~\bibnamefont{Tandia}},
  \bibinfo{author}{\bibfnamefont{K.~D.} \bibnamefont{Vargheese}},
  \bibinfo{author}{\bibfnamefont{J.~C.} \bibnamefont{Mauro}}, \bibnamefont{and}
  \bibinfo{author}{\bibfnamefont{A.~K.} \bibnamefont{Varshneya}},
  \bibinfo{journal}{Journal of Non-Crystalline Solids}
  \textbf{\bibinfo{volume}{358}}, \bibinfo{pages}{316} (\bibinfo{year}{2012}).

\bibitem[{\citenamefont{Plimpton}(1995)}]{plimpton_fast_1995}
\bibinfo{author}{\bibfnamefont{S.}~\bibnamefont{Plimpton}},
  \bibinfo{journal}{Journal of computational physics}
  \textbf{\bibinfo{volume}{117}}, \bibinfo{pages}{1} (\bibinfo{year}{1995}).

\bibitem[{\citenamefont{Cormack et~al.}(2002)\citenamefont{Cormack, Du, and
  Zeitler}}]{cormack_alkali_2002}
\bibinfo{author}{\bibfnamefont{A.~N.} \bibnamefont{Cormack}},
  \bibinfo{author}{\bibfnamefont{J.}~\bibnamefont{Du}}, \bibnamefont{and}
  \bibinfo{author}{\bibfnamefont{T.~R.} \bibnamefont{Zeitler}},
  \bibinfo{journal}{Phys. Chem. Chem. Phys.} \textbf{\bibinfo{volume}{4}},
  \bibinfo{pages}{3193} (\bibinfo{year}{2002}).

\bibitem[{\citenamefont{Bauchy}(2012)}]{bauchy_structural_2012}
\bibinfo{author}{\bibfnamefont{M.}~\bibnamefont{Bauchy}}, \bibinfo{journal}{The
  Journal of Chemical Physics} \textbf{\bibinfo{volume}{137}},
  \bibinfo{pages}{044510} (\bibinfo{year}{2012}).

\bibitem[{\citenamefont{Bauchy et~al.}(2013)\citenamefont{Bauchy, Guillot,
  Micoulaut, and Sator}}]{bauchy_viscosity_2013}
\bibinfo{author}{\bibfnamefont{M.}~\bibnamefont{Bauchy}},
  \bibinfo{author}{\bibfnamefont{B.}~\bibnamefont{Guillot}},
  \bibinfo{author}{\bibfnamefont{M.}~\bibnamefont{Micoulaut}},
  \bibnamefont{and} \bibinfo{author}{\bibfnamefont{N.}~\bibnamefont{Sator}},
  \bibinfo{journal}{Chemical Geology} \textbf{\bibinfo{volume}{346}},
  \bibinfo{pages}{47} (\bibinfo{year}{2013}).

\bibitem[{\citenamefont{Xiang et~al.}(2013)\citenamefont{Xiang, Du, Smedskjaer,
  and Mauro}}]{xiang_structure_2013}
\bibinfo{author}{\bibfnamefont{Y.}~\bibnamefont{Xiang}},
  \bibinfo{author}{\bibfnamefont{J.}~\bibnamefont{Du}},
  \bibinfo{author}{\bibfnamefont{M.~M.} \bibnamefont{Smedskjaer}},
  \bibnamefont{and} \bibinfo{author}{\bibfnamefont{J.~C.} \bibnamefont{Mauro}},
  \bibinfo{journal}{The Journal of Chemical Physics}
  \textbf{\bibinfo{volume}{139}}, \bibinfo{pages}{044507}
  (\bibinfo{year}{2013}).

\bibitem[{\citenamefont{Barkema and Mousseau}(1996)}]{barkema_event-based_1996}
\bibinfo{author}{\bibfnamefont{G.~T.} \bibnamefont{Barkema}} \bibnamefont{and}
  \bibinfo{author}{\bibfnamefont{N.}~\bibnamefont{Mousseau}},
  \bibinfo{journal}{Physical Review Letters} \textbf{\bibinfo{volume}{77}},
  \bibinfo{pages}{4358} (\bibinfo{year}{1996}).

\bibitem[{\citenamefont{Richard et~al.}(2005)\citenamefont{Richard, Nicodemi,
  Delannay, Ribi{\`e}re, and Bideau}}]{richard_slow_2005}
\bibinfo{author}{\bibfnamefont{P.}~\bibnamefont{Richard}},
  \bibinfo{author}{\bibfnamefont{M.}~\bibnamefont{Nicodemi}},
  \bibinfo{author}{\bibfnamefont{R.}~\bibnamefont{Delannay}},
  \bibinfo{author}{\bibfnamefont{P.}~\bibnamefont{Ribi{\`e}re}},
  \bibnamefont{and} \bibinfo{author}{\bibfnamefont{D.}~\bibnamefont{Bideau}},
  \bibinfo{journal}{Nature Materials} \textbf{\bibinfo{volume}{4}},
  \bibinfo{pages}{121} (\bibinfo{year}{2005}).

\bibitem[{\citenamefont{M{\"o}bius and
  Heussinger}(2014)}]{mobius_irreversibility_2014}
\bibinfo{author}{\bibfnamefont{R.}~\bibnamefont{M{\"o}bius}} \bibnamefont{and}
  \bibinfo{author}{\bibfnamefont{C.}~\bibnamefont{Heussinger}},
  \bibinfo{journal}{Soft Matter} \textbf{\bibinfo{volume}{10}},
  \bibinfo{pages}{4806} (\bibinfo{year}{2014}).

\bibitem[{\citenamefont{Lacks}(2001)}]{lacks_energy_2001}
\bibinfo{author}{\bibfnamefont{D.~J.} \bibnamefont{Lacks}},
  \bibinfo{journal}{Physical Review Letters} \textbf{\bibinfo{volume}{87}},
  \bibinfo{pages}{225502} (\bibinfo{year}{2001}).

\bibitem[{\citenamefont{Lacks and Osborne}(2004)}]{lacks_energy_2004}
\bibinfo{author}{\bibfnamefont{D.~J.} \bibnamefont{Lacks}} \bibnamefont{and}
  \bibinfo{author}{\bibfnamefont{M.~J.} \bibnamefont{Osborne}},
  \bibinfo{journal}{Physical Review Letters} \textbf{\bibinfo{volume}{93}},
  \bibinfo{pages}{255501} (\bibinfo{year}{2004}).

\bibitem[{\citenamefont{Utz et~al.}(2000)\citenamefont{Utz, Debenedetti, and
  Stillinger}}]{utz_atomistic_2000}
\bibinfo{author}{\bibfnamefont{M.}~\bibnamefont{Utz}},
  \bibinfo{author}{\bibfnamefont{P.~G.} \bibnamefont{Debenedetti}},
  \bibnamefont{and} \bibinfo{author}{\bibfnamefont{F.~H.}
  \bibnamefont{Stillinger}}, \bibinfo{journal}{Physical Review Letters}
  \textbf{\bibinfo{volume}{84}}, \bibinfo{pages}{1471} (\bibinfo{year}{2000}).

\bibitem[{\citenamefont{Lyulin and Michels}(2007)}]{lyulin_time_2007}
\bibinfo{author}{\bibfnamefont{A.~V.} \bibnamefont{Lyulin}} \bibnamefont{and}
  \bibinfo{author}{\bibfnamefont{M.~A.~J.} \bibnamefont{Michels}},
  \bibinfo{journal}{Physical Review Letters} \textbf{\bibinfo{volume}{99}},
  \bibinfo{pages}{085504} (\bibinfo{year}{2007}).

\bibitem[{\citenamefont{Rottler and Warren}(2008)}]{rottler_deformation_2008}
\bibinfo{author}{\bibfnamefont{J.}~\bibnamefont{Rottler}} \bibnamefont{and}
  \bibinfo{author}{\bibfnamefont{M.}~\bibnamefont{Warren}},
  \bibinfo{journal}{The European Physical Journal Special Topics}
  \textbf{\bibinfo{volume}{161}}, \bibinfo{pages}{55} (\bibinfo{year}{2008}).

\bibitem[{\citenamefont{Bonn et~al.}(2002)\citenamefont{Bonn, Tanase, Abou,
  Tanaka, and Meunier}}]{bonn_laponite:_2002}
\bibinfo{author}{\bibfnamefont{D.}~\bibnamefont{Bonn}},
  \bibinfo{author}{\bibfnamefont{S.}~\bibnamefont{Tanase}},
  \bibinfo{author}{\bibfnamefont{B.}~\bibnamefont{Abou}},
  \bibinfo{author}{\bibfnamefont{H.}~\bibnamefont{Tanaka}}, \bibnamefont{and}
  \bibinfo{author}{\bibfnamefont{J.}~\bibnamefont{Meunier}},
  \bibinfo{journal}{Physical Review Letters} \textbf{\bibinfo{volume}{89}},
  \bibinfo{pages}{015701} (\bibinfo{year}{2002}).

\bibitem[{\citenamefont{Viasnoff and
  Lequeux}(2002)}]{viasnoff_rejuvenation_2002}
\bibinfo{author}{\bibfnamefont{V.}~\bibnamefont{Viasnoff}} \bibnamefont{and}
  \bibinfo{author}{\bibfnamefont{F.}~\bibnamefont{Lequeux}},
  \bibinfo{journal}{Physical Review Letters} \textbf{\bibinfo{volume}{89}},
  \bibinfo{pages}{065701} (\bibinfo{year}{2002}).

\bibitem[{\citenamefont{Priezjev}(2013)}]{priezjev_heterogeneous_2013}
\bibinfo{author}{\bibfnamefont{N.~V.} \bibnamefont{Priezjev}},
  \bibinfo{journal}{Physical Review E} \textbf{\bibinfo{volume}{87}},
  \bibinfo{pages}{052302} (\bibinfo{year}{2013}).

\bibitem[{\citenamefont{Fiocco et~al.}(2013)\citenamefont{Fiocco, Foffi, and
  Sastry}}]{fiocco_oscillatory_2013}
\bibinfo{author}{\bibfnamefont{D.}~\bibnamefont{Fiocco}},
  \bibinfo{author}{\bibfnamefont{G.}~\bibnamefont{Foffi}}, \bibnamefont{and}
  \bibinfo{author}{\bibfnamefont{S.}~\bibnamefont{Sastry}},
  \bibinfo{journal}{Physical Review E} \textbf{\bibinfo{volume}{88}},
  \bibinfo{pages}{020301} (\bibinfo{year}{2013}).

\bibitem[{\citenamefont{Parrinello and
  Rahman}(1981)}]{parrinello_polymorphic_1981}
\bibinfo{author}{\bibfnamefont{M.}~\bibnamefont{Parrinello}} \bibnamefont{and}
  \bibinfo{author}{\bibfnamefont{A.}~\bibnamefont{Rahman}},
  \bibinfo{journal}{Journal of Applied Physics} \textbf{\bibinfo{volume}{52}},
  \bibinfo{pages}{7182} (\bibinfo{year}{1981}).

\bibitem[{\citenamefont{Vineyard}(1957)}]{vineyard_frequency_1957}
\bibinfo{author}{\bibfnamefont{G.~H.} \bibnamefont{Vineyard}},
  \bibinfo{journal}{Journal of Physics and Chemistry of Solids}
  \textbf{\bibinfo{volume}{3}}, \bibinfo{pages}{121} (\bibinfo{year}{1957}).

\bibitem[{\citenamefont{Masoero et~al.}(2013)\citenamefont{Masoero, Manzano,
  Del~Gado, Pellenq, Ulm, and Yip}}]{masoero_kinetic_2013}
\bibinfo{author}{\bibfnamefont{E.}~\bibnamefont{Masoero}},
  \bibinfo{author}{\bibfnamefont{H.}~\bibnamefont{Manzano}},
  \bibinfo{author}{\bibfnamefont{E.}~\bibnamefont{Del~Gado}},
  \bibinfo{author}{\bibfnamefont{R.~J.~M.} \bibnamefont{Pellenq}},
  \bibinfo{author}{\bibfnamefont{F.~J.} \bibnamefont{Ulm}}, \bibnamefont{and}
  \bibinfo{author}{\bibfnamefont{S.}~\bibnamefont{Yip}},
  \bibinfo{journal}{Mechanics and Physics of Creep, Shrinkage, and Durability
  of Concrete: A Tribute to Zdenk P. Bazant} pp. \bibinfo{pages}{166--173}
  (\bibinfo{year}{2013}).

\bibitem[{\citenamefont{Potuzak et~al.}(2011)\citenamefont{Potuzak, Welch, and
  Mauro}}]{potuzak_topological_2011}
\bibinfo{author}{\bibfnamefont{M.}~\bibnamefont{Potuzak}},
  \bibinfo{author}{\bibfnamefont{R.~C.} \bibnamefont{Welch}}, \bibnamefont{and}
  \bibinfo{author}{\bibfnamefont{J.~C.} \bibnamefont{Mauro}},
  \bibinfo{journal}{The Journal of Chemical Physics}
  \textbf{\bibinfo{volume}{135}}, \bibinfo{pages}{214502}
  (\bibinfo{year}{2011}).

\bibitem[{\citenamefont{Simdyankin and
  Mousseau}(2003)}]{simdyankin_relationship_2003}
\bibinfo{author}{\bibfnamefont{S.~I.} \bibnamefont{Simdyankin}}
  \bibnamefont{and} \bibinfo{author}{\bibfnamefont{N.}~\bibnamefont{Mousseau}},
  \bibinfo{journal}{Physical Review E} \textbf{\bibinfo{volume}{68}},
  \bibinfo{pages}{041110} (\bibinfo{year}{2003}).

\bibitem[{\citenamefont{Mauro and
  Smedskjaer}(2012{\natexlab{a}})}]{mauro_unified_2012}
\bibinfo{author}{\bibfnamefont{J.~C.} \bibnamefont{Mauro}} \bibnamefont{and}
  \bibinfo{author}{\bibfnamefont{M.~M.} \bibnamefont{Smedskjaer}},
  \bibinfo{journal}{Physica A: Statistical Mechanics and its Applications}
  \textbf{\bibinfo{volume}{391}}, \bibinfo{pages}{6121}
  (\bibinfo{year}{2012}{\natexlab{a}}).

\bibitem[{\citenamefont{Mauro and
  Smedskjaer}(2012{\natexlab{b}})}]{mauro_minimalist_2012}
\bibinfo{author}{\bibfnamefont{J.~C.} \bibnamefont{Mauro}} \bibnamefont{and}
  \bibinfo{author}{\bibfnamefont{M.~M.} \bibnamefont{Smedskjaer}},
  \bibinfo{journal}{Physica A: Statistical Mechanics and its Applications}
  \textbf{\bibinfo{volume}{391}}, \bibinfo{pages}{3446}
  (\bibinfo{year}{2012}{\natexlab{b}}).

\bibitem[{\citenamefont{Bauchy and Micoulaut}(2015)}]{bauchy_densified_2015}
\bibinfo{author}{\bibfnamefont{M.}~\bibnamefont{Bauchy}} \bibnamefont{and}
  \bibinfo{author}{\bibfnamefont{M.}~\bibnamefont{Micoulaut}},
  \bibinfo{journal}{Nature Communications} \textbf{\bibinfo{volume}{6}}
  (\bibinfo{year}{2015}).

\bibitem[{\citenamefont{Bauchy and Micoulaut}(2011)}]{bauchy_pockets_2011}
\bibinfo{author}{\bibfnamefont{M.}~\bibnamefont{Bauchy}} \bibnamefont{and}
  \bibinfo{author}{\bibfnamefont{M.}~\bibnamefont{Micoulaut}},
  \bibinfo{journal}{Physical Review B} \textbf{\bibinfo{volume}{83}},
  \bibinfo{pages}{184118} (\bibinfo{year}{2011}).

\bibitem[{\citenamefont{Abdolhosseini~Qomi
  et~al.}(2014)\citenamefont{Abdolhosseini~Qomi, Bauchy, Ulm, and
  Pellenq}}]{abdolhosseini_qomi_anomalous_2014}
\bibinfo{author}{\bibfnamefont{M.~J.} \bibnamefont{Abdolhosseini~Qomi}},
  \bibinfo{author}{\bibfnamefont{M.}~\bibnamefont{Bauchy}},
  \bibinfo{author}{\bibfnamefont{F.-J.} \bibnamefont{Ulm}}, \bibnamefont{and}
  \bibinfo{author}{\bibfnamefont{R.~J.-M.} \bibnamefont{Pellenq}},
  \bibinfo{journal}{The Journal of Chemical Physics}
  \textbf{\bibinfo{volume}{140}}, \bibinfo{pages}{054515}
  (\bibinfo{year}{2014}).

\bibitem[{\citenamefont{Mousseau}(2000)}]{mousseau_cooperative_2000}
\bibinfo{author}{\bibfnamefont{N.}~\bibnamefont{Mousseau}},
  \bibinfo{journal}{arXiv:cond-mat/0004356}  (\bibinfo{year}{2000}).

\bibitem[{\citenamefont{Chakravarty et~al.}(2005)\citenamefont{Chakravarty,
  Georgiev, Boolchand, and Micoulaut}}]{chakravarty_ageing_2005}
\bibinfo{author}{\bibfnamefont{S.}~\bibnamefont{Chakravarty}},
  \bibinfo{author}{\bibfnamefont{D.~G.} \bibnamefont{Georgiev}},
  \bibinfo{author}{\bibfnamefont{P.}~\bibnamefont{Boolchand}},
  \bibnamefont{and}
  \bibinfo{author}{\bibfnamefont{M.}~\bibnamefont{Micoulaut}},
  \bibinfo{journal}{Journal of Physics: Condensed Matter}
  \textbf{\bibinfo{volume}{17}}, \bibinfo{pages}{L1} (\bibinfo{year}{2005}).

\bibitem[{\citenamefont{Fiocco et~al.}(2014)\citenamefont{Fiocco, Foffi, and
  Sastry}}]{fiocco_encoding_2014}
\bibinfo{author}{\bibfnamefont{D.}~\bibnamefont{Fiocco}},
  \bibinfo{author}{\bibfnamefont{G.}~\bibnamefont{Foffi}}, \bibnamefont{and}
  \bibinfo{author}{\bibfnamefont{S.}~\bibnamefont{Sastry}},
  \bibinfo{journal}{Physical Review Letters} \textbf{\bibinfo{volume}{112}},
  \bibinfo{pages}{025702} (\bibinfo{year}{2014}).

\bibitem[{\citenamefont{Fiocco et~al.}(2015)\citenamefont{Fiocco, Foffi, and
  Sastry}}]{fiocco_memory_2015}
\bibinfo{author}{\bibfnamefont{D.}~\bibnamefont{Fiocco}},
  \bibinfo{author}{\bibfnamefont{G.}~\bibnamefont{Foffi}}, \bibnamefont{and}
  \bibinfo{author}{\bibfnamefont{S.}~\bibnamefont{Sastry}},
  \bibinfo{journal}{arXiv:1503.00478 [cond-mat]}  (\bibinfo{year}{2015}).

\bibitem[{\citenamefont{Mousseau et~al.}(2001)\citenamefont{Mousseau,
  Derreumaux, Barkema, and Malek}}]{mousseau_sampling_2001}
\bibinfo{author}{\bibfnamefont{N.}~\bibnamefont{Mousseau}},
  \bibinfo{author}{\bibfnamefont{P.}~\bibnamefont{Derreumaux}},
  \bibinfo{author}{\bibfnamefont{G.~T.} \bibnamefont{Barkema}},
  \bibnamefont{and} \bibinfo{author}{\bibfnamefont{R.}~\bibnamefont{Malek}},
  \bibinfo{journal}{Journal of Molecular Graphics and Modelling}
  \textbf{\bibinfo{volume}{19}}, \bibinfo{pages}{78} (\bibinfo{year}{2001}).

\bibitem[{\citenamefont{Phillips}(2009)}]{phillips_scaling_2009}
\bibinfo{author}{\bibfnamefont{J.~C.} \bibnamefont{Phillips}},
  \bibinfo{journal}{Physical Review E} \textbf{\bibinfo{volume}{80}}
  (\bibinfo{year}{2009}).

\end{thebibliography}

\end{document}